\documentclass{PoS}

\newcommand{\be}{\begin{equation}}
\newcommand{\en}{\end{equation}}
\newcommand{\bea}{\begin{eqnarray}}
\newcommand{\ena}{\end{eqnarray}}
\newcommand{\dlangle}{\left\langle \kern-.17em \left\langle}
\newcommand{\drangle}{\right\rangle \kern-.17em \right\rangle}
\newcommand{\hbo}{\hbox to 1 true cm {\hfill } }

\newcommand{\e}{\mathrm{e}}
\newcommand{\lb}{\langle \kern-.17em \langle} 
\newcommand{\rb}{\rangle \kern-.17em \rangle }
\newcommand{\DE}{\delta E}
\newcommand{\dd}{d}

\title{Density of states}

\ShortTitle{DoS}

\author{\speaker{Kurt Langfeld}\thanks{
This work is supported in part by the Leverhulme Trust (grant
RPG-2014-118) and STFC (grant ST/L000350/1).} \\
        Department for the Mathematical Sciences, University of Liverpool, 
        Liverpool, L69 7ZL, UK \\ 
        E-mail: \email{kurt.langfeld@liverpool.ac.uk}}

\abstract{
  Although Monte Carlo calculations using Importance Sampling have
  matured into the most widely employed method for determining first
  principle results in QCD, they spectacularly fail for theories with
  a sign problem or for which certain rare configurations play an
  important role. Non-Markovian Random walks, based upon iterative
  refinements of the density-of-states, overcome such overlap
  problems. I will review the Linear Logarithmic
  Relaxation (LLR) method  and, in particular, focus onto ergodicity
  and exponential error suppression. Applications include the
  high-state Potts model, SU(2) and SU(3) Yang-Mills theories as
  well as a quantum field theory with a strong sign problem: QCD at finite
  densities of heavy quarks.  }  

\FullConference{34th annual International Symposium on Lattice Field Theory\\
		24-30 July 2016\\
		University of Southampton, UK}

\begin{document}

\section{Introduction}

Markov chain Monte-Carlo (MCMC) simulations of quantum field theories
discretised on a space-time lattice have matured into a powerful
quantitative tool for ab-initio calculations. Among other
non-perturbative techniques, the almost unique characteristic is the
control over error margins within the approach itself.

\medskip 
Still nowadays, Markov chain Monte-Carlo simulations face two severe
limitations 

(i) due to ergodicity problems

(ii) in cases of a non-positive Gibbs factor.

Both issues are far from technical problems only, but are
intrinsically connected to physics phenomenons that entirely evade
rigorous results from MCMC.

\medskip 
Let me just mention two examples with
class (i) issues: In cases of pure Yang-Mills theories with
configurations falling into sectors of the so-called topological
charge, even advanced methods (such as Hybrid Monte Carlo type
algorithms) fail to sweep between sectors on large volumes
(see~\cite{Schaefer:2010hu} for a precise definition and an analysis of
the issue). The second example concerns the wide range of thermal or
quantum field theories with a first order phase transition: at
criticality, the configuration space falls into distinct sets of
configurations that are equally relevant. For the transfer between
sets, MCMC needs either theory dependent information for global updates
(cluster algorithms fall into this class) or needs to fall back to
rather local update strategies. In the later case, configurations need
to pass through regions of low probability. In fact, this probability
is exponentially small with the volume $V$, i.e., $\exp \{ - \sigma
V\}$\footnote{$\sigma $ is the tension of interfaces between pockets
  of different phase material at criticality.} , and ergodicity is
hampered at the practical level for relevant sized systems.  

\medskip
In class (ii), we e.g.~find quantum field theories (QFTs) with a topological
$\theta $-term or, of even bigger phenomenological relevance, QFTs
with matter at finite densities or, more precisely, at finite chemical
potentials. Different strategies have been explored over the decades,
and I refer to the reviews at the annual Lattice Conference for a
summary of recent progress \cite{Borsanyi:2015axp,
  Sexty:2014dxa,Gattringer:2014nxa,Aarts:2013lcm,Wolff:2010zu,deForcrand:2010ys,Chandrasekharan:2008gp}. 

\medskip
A more radical approach is to abandon MCMC simulations altogether and
to explore instead {\it Non-Markovian Random Walk} simulations. These do not
rely on an update of the configurations according to Importance
Sampling with respect to a positive Gibbs factor, which makes them
well suited to study class (ii) theories. One particular set-up that
is particularly relevant for QFTs with class (i) and (ii) issues uses
the inverse density-of-states~\footnote{A precise definition will be
  given below.} as a measure for updating
configuration. This measure is (semi-) positive definite by
definition, and aims to generate a random walk in configuration
space even in ``deprived'' regions with very low probabilistic measure.
To my knowledge, the first of its kind is the Multicanonical Algorithm
by Berg and Neuhaus~\cite{Berg:1992qua}. The same re-weighting approach
is also at the heart of the Wang-Landau approach~\cite{Wang:2001ab}
or, more recently a version adapted to continuous QFT degrees of
freedom, the LLR method~\cite{Langfeld:2012ah,Langfeld:2015fua}, which
will be the focal point of this paper (also
see~\cite{Gattringer:2016kco} for a recent review).  

\medskip
\begin{figure}
  \includegraphics[height=7cm]{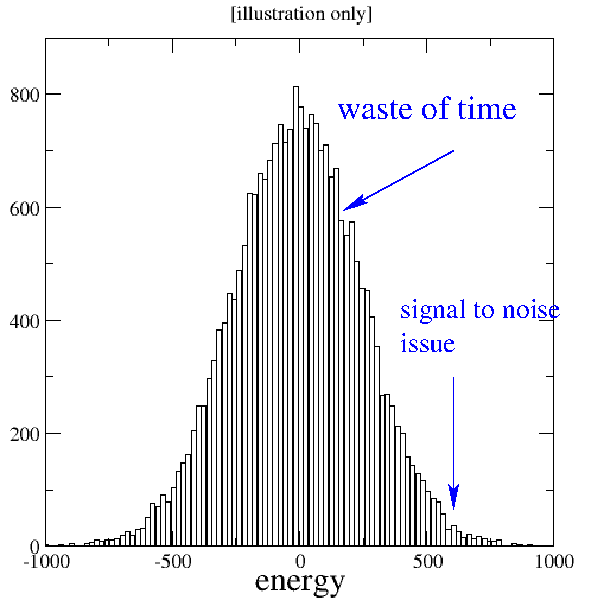}
  \hspace{0.5cm} 
  \includegraphics[height=7cm]{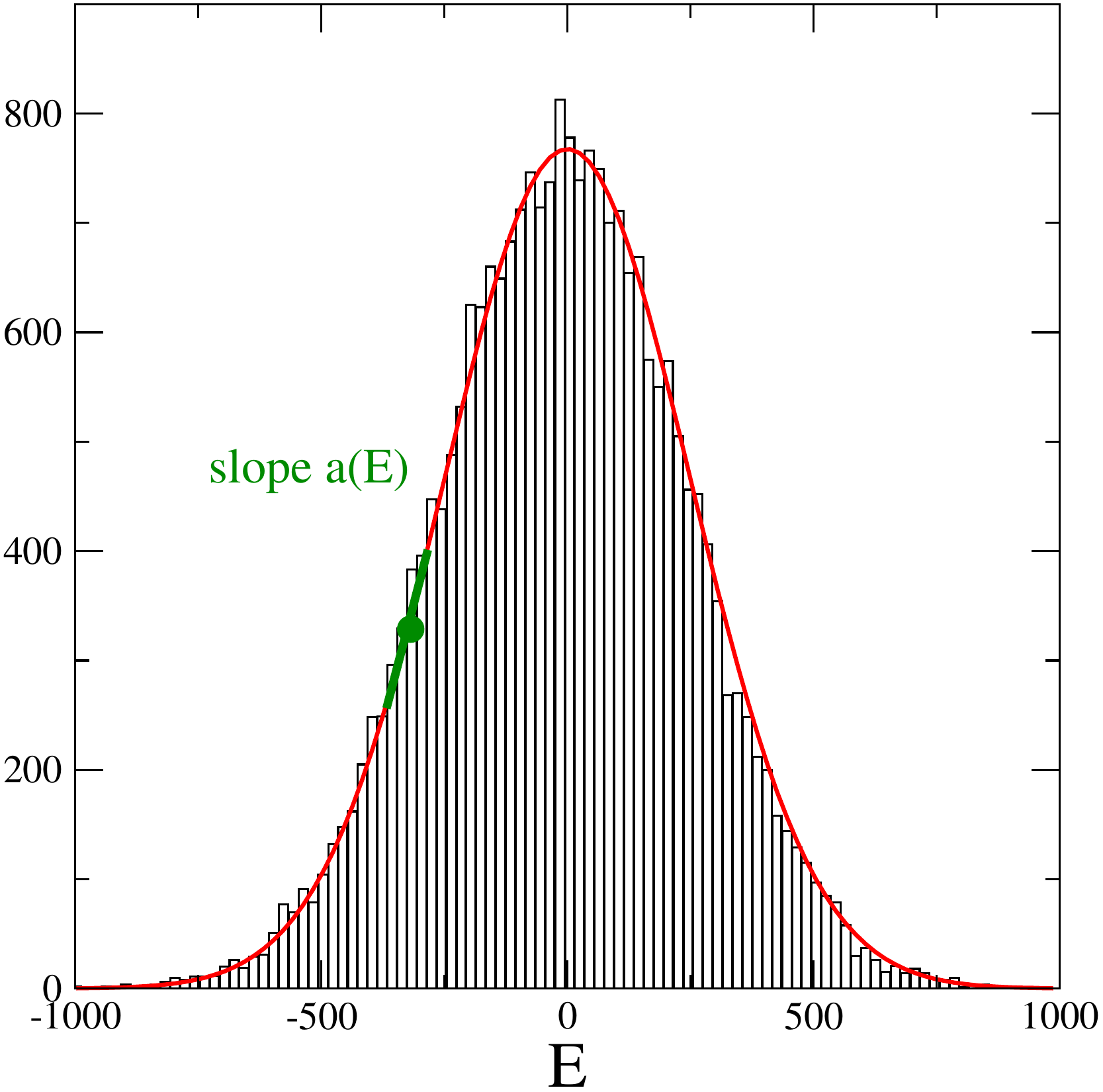}
  \caption{\label{fig:1} Illustration of the density of states $\rho
    (E)$ }
\end{figure}
Let us now look at the definition of the density-of-states $\rho (E)$
in a QFT setting with degrees of freedom $\phi (x)$. In its simplest
form, it quantifies the number of configurations for a given action
hyper-surface in configuration space:
\be 
\rho(E) = \int {\cal D} \phi \;  \delta \Bigl( S[\phi]-E \Bigr) \, ,
\label{eq:1}
\en
The probabilistic weight $P(E)$ is given by the product of density and
Gibbs factor, \break $P(E) = \rho (E) \, \exp \{ \beta E \} $, and the
partition function is recovered by performing a one-dimensional
integral
\be 
Z(\beta)=\int {\cal D} \phi \;  \e^{\beta S[\phi]} \, = \, \int \dd E
\; P(E) \; = \; \int \dd E \; \rho(E) \; \e^{\beta E} \, , 
\label{eq:2}
\en
as can be easily verified by inserting (\ref{eq:1}) into
(\ref{eq:2}). 
The density of states (\ref{eq:1}) could be straightforwardly
estimated by generating configurations $\{\phi \}$ according to the
plain measure, calculate the corresponding action $S[\phi]$ and
accumulate statistics in a histogram. The result could look like the
histogram in figure~\ref{fig:1}, left panel. In order to obtain the
statistical weight $P(E)$ with which the configuration contributes to
the partition function $Z$, the density $\rho (E)$ is multiplied by
the Gibbs factor. This exponentially amplifies the relevance of the
configurations with large action $E$. Note, however, that we have
relatively few events in the histogram at relevant values for $E$. The
majority is generated around $E \approx 0$, which only marginally
contribute to $Z$. Hence, we would end up with a poor signal for this
amount of statistics. The target of the density-of-states method (in
the LLR formulation) is to estimate the slope $a(E)$ at any chosen
point $E$ (see figure~\ref{fig:1}, right panel):
\be
a(E) \; = \; \frac{ d \, \ln \rho(E) }{ dE } \; .
\label{eq:3}
\en
Once this slope $a(E)$ is obtained, the density $\rho (E)$ can be
reconstructed up to a multiplicative factor by integration:
\be 
\rho(E) \; = \; \rho (E_0) \; \exp \left\{ \int _{E_0}^E a(E^\prime
) \; dE^\prime \right\} \; . 
\label{eq:3b}
\en 
The result of this integration is indicated by the red line in
figure~\ref{fig:1}, right panel. As detailed below, crucial is that
the estimate for $a(E)$ can be obtained with roughly the same
statistical error independent of $E$ and even if $E$ is from a region
of overall low weight $P(E)$. This is how the LLR approach circumvents
overlap problems.

\section{The LLR method}

\subsection{Finding the LLR coefficients }

While the Multicanonical approach~\cite{Berg:1992qua} and the
Wang-Landau method~\cite{Wang:2001ab} directly target the density
$\rho (E)$, the coefficients $a(E)$ (\ref{eq:3}) are the primary
target of the LLR method~\cite{Langfeld:2012ah}. Key ingredient is the
Monte-Carlo expectation value
\bea 
\dlangle W[\phi] \drangle_{E} (a)
&=& \frac{1}{{\cal N}_E} \int {\cal D} \phi \; \Omega
_{[E,\DE]}(S[\phi]) \; W[\phi] \; \,  \e^{-a S[\phi] } \; .
\label{eq:4} \\
{\cal N}_E &=& \int {\cal D} \phi \; \Omega _{[E,\DE]} \; \, 
\e^{-a S[\phi] } \; , 
\label{eq:5} 
\ena
where $a$ is, for the moment, an external parameter, $W[\phi]$ is an
observable, and $\Omega _{[E,\DE]} $ is the so-called Window
function. This is the only place where the $E$ dependence of $\dlangle
W[\phi] \drangle_{E} $ comes into play. Note that the double
expectation value $\dlangle \ldots \drangle_{E} $ appears as a
standard MC average and can be estimated with standard Importance
Sampling methods. Historically and to maintain a
close connection to the Wang-Landau method, the Window function
restricts the simulation to an interval of size $\DE $ around $E$:
\be 
\Omega _{[E,\DE]} (S) \; = \; \left\{ \begin{array}{ll} 
1 & \hbox{for} \; \; \; E-\frac{\DE}{2} \leq S \leq E + \frac{\DE}{2} \\ 
0 & \hbox{otherwise . } \end{array} \right. 
\label{eq:6}
\en 
The choice of a {\it smooth} window function is of great importance if
advanced MC simulation techniques such as the Hybrid Monte Carlo
method are employed. In this case, the window function contributes to
the MC drift force, and the algorithm enjoys the familiar high
acceptance rates for configurations. Below, I will show results
obtained with the choice
\be 
\Omega _{[E,\DE]} (S) \; = \; \exp \left\{ - \, \frac{ (S-E)^2 }{ \DE
  ^2 } \right\} \; . 
\label{eq:7}
\en
We will be interested in the limit $\DE \to 0$. As will become clear
below, the only important constraint for potential Window functions is
that $\Omega (S)$ is symmetric around $S=E$ and that it decreases
sufficiently fast for $S \to \pm \infty $. 

\medskip
We are now prepared to study the LLR method to find the slope $a(E)$
in (\ref{eq:3}). To this aim, we choose $W[\phi]=S[\phi]- E$ and use the
definition of $\rho $ [using $E^\prime$ as integration variable to
  avoid a clash of notation] (\ref{eq:1}) in (\ref{eq:4},\ref{eq:5}),
which become:
\bea 
\dlangle W[\phi] \drangle_{E} (a)
&=& \frac{1}{{\cal N}_E} \int dE ^\prime \; \Omega
_{[E,\DE]}(E^\prime ) \; \Bigl(E^\prime -E \Bigr) \; \,
\rho(E^\prime) \, \e^{-a E^\prime } \; . 
\label{eq:8} \\
{\cal N}_E &=& \int dE \; \Omega _{[E,\DE]} (E^\prime) \; \, 
\e^{-a E^\prime } \; , 
\label{eq:9} 
\ena
The Window function confines the integration to a small, $\DE$ sized
interval around $E$. For sufficiently smooth densities $\rho
(E^\prime)$, we might expand in leading order 
\be
\ln \rho (E^\prime ) \; \approx \; \ln \rho (E) + \frac{ d \, \ln \rho (E)
}{dE} \; \Bigl( E^\prime - E \Bigr) \; = \; \ln \rho (E) + a(E) \;
\Bigl( E^\prime - E \Bigr) \; . 
\label{eq:10} 
\en
Using (\ref{eq:10}) in (\ref{eq:8},\ref{eq:9}), we find
\bea 
\dlangle S[\phi] - E \drangle_{E} (a)
&\approx & \frac{1}{N_E} \int dE ^\prime \; \Omega
_{[E,\DE]}(E^\prime ) \; \Bigl(E^\prime -E \Bigr) \; \,
 \exp \Bigl\{ - [a(E) - a] ( E^\prime -E) \Bigr\} \; , 
\nonumber \\ 
{N}_E &=& \int dE \; \Omega _{[E,\DE]} (E^\prime) \; \, 
 \exp \Bigl\{ - [a(E) - a] ( E^\prime -E) \Bigr\} \; .
\nonumber 
\ena
Since $\Omega _{[E,\DE]} (E^\prime)$ is symmetric around $E$, we find
for sufficiently small $\DE$ the key equation:
\be
\lim _{\DE \to 0} \dlangle S[\phi] - E \drangle_{E} \Bigl( a= a(E)
\Bigr) \; = \; 0 \; .
\label{eq:15}
\en
Remember that the expectation value $\dlangle S[\phi] - E
\drangle_{E}(a)$ is accessible by standard MC simulations for any
external parameter $a$. Hence, the LLR coefficient $a(E)$ can be
obtained by solving the {\it stochastic non-linear equation}: 
\be 
\dlangle S[\phi] - E \drangle_{E} ( a ) \; = \; 0 \; .
\label{eq:16}
\en
For finite $\DE$, it can be shown~\cite{Langfeld:2015fua} that
$\dlangle S[\phi] - E \drangle_{E} (a ) $ is a monotonic
function and, thus, the solution is unique. A thorough study of the
truncation error yields for the solution~\cite{Langfeld:2015fua}:
\be 
\dlangle \Delta S[\phi] -E \drangle_E (a) \; = \; 0 \hbo \Leftrightarrow \hbo 
a \; = \; a(E) \; + \; {\cal O}\Bigl(\DE^2\Bigr) \; .
\label{eq:17}
\en
For an alternative method to calculate the coefficients $a(E)$, we
point the reader to the Functional Fit
Approach~\cite{Gattringer:2015lra,Giuliani:2016tlu}.

\subsection{Features of the LLR algorithm}

How do we solve the stochastic equation (\ref{eq:16}) in practice and
how do we control errors for observables?

\medskip
In practice, only MC estimators are available for the expectation value
$\dlangle \Delta S[\phi] -E \drangle_E (a) $. Let us assume that $M$
MC samples are used to estimate the expectation value. Any numerical
procedure for solving the stochastic equation will produce a
statistical ensemble $\{ a \}_M$ of approximate solutions. For
practicability and robustness of the density-of-states method, we need
to achieve two goals:

(1) The average over the ensemble $\{ a \}_M$ needs to agree with the
true solution: $ \langle a \rangle = a(E)$.

(2) For sufficiently large $M$, the distribution for $\{ a \}_M$
should be normal.

Both properties would provide us with the capability to control
errors: For each $E$ and for a large enough choice $M$, we would
generate $n_B$ approximate solutions $\bar{a}^{(M)}_n$, $n=1 \ldots
n_B$. Because of property (2), the $n_b$ copies of potential solutions
can be used in a standard bootstrap analysis for derived observables.

\medskip 
The LLR algorithms comes with a specific proposal to accomplish the
above two goals. Starting point is the fix-point iteration
\be
a^{(n+1)} \; = \; a^{(n)} \; + \;  c_n \; \frac{12 }{\DE ^2 }  \; 
\dlangle S[\phi ] - E  \drangle_E   (a^{(n)}) \; ,
 \label{eq:20}
 \en
with coefficients $c_n$ yet to be specified. For $c_n=1$, the above
iteration is inspired by the Newton-Raphson method. The issue is that
there a two sources of error: the convergence error of the iteration
and the statistical uncertainty induced by replacing the expectation
value $\dlangle S[\phi ] - E  \drangle_E$ by MC estimators. The
mathematical framework to solve this problem has been already provided
by Robbins and Monro in the fifties~\cite{Robbins1951}: if the 
coefficients satisfy
$$
\sum^{\infty}_{n=0}c_n = \infty \hbo \mbox{ and } \hbo 
\sum^{\infty}_{n=0}c^2_n < \infty \, . 
$$
and if the iteration is truncated at some (large)
$n=n_{RM}$, the corresponding values $a^{(n_{RM})}$ are normal distributed
with the mean coinciding with the true solution. Moreover, the
limiting case, i.e., $c_n=1/(n+1)$, is optimal~\cite{Robbins1951}. 

\medskip 
Other methods such as the Functional Fit
Approach~\cite{Gattringer:2015lra,Giuliani:2016tlu} have been proposed
to extract the slope $a(E)$ although more work to establish 
a systematic error control might be needed. 

\medskip 
Once the LLR coefficients $a(E)$ have been obtained for a set of values $E$, 
the density $\rho (E)$ is reconstructed by integration (see
(\ref{eq:3b})). This integration is necessarily approximate, and we
call $\widetilde{\rho}(E)$ the approximation to the density of
states. In practice, $\widetilde{\rho}(E)$ is affected by truncation
errors, i.e., using a finite non-zero value for $\DE$, and statistical
errors for each value $E$ of the set. Key observation is, however,
that the LLR approach features an {\it exponential error suppression}
in the sense that~\cite{Langfeld:2015fua}
\be
\rho (E) \; = \; \widetilde{\rho}(E) \; \exp \Bigl\{
\hbox{truncation \& statistical error} \Bigr\} \; , 
\label{eq:21}
\en
i.e., the {\it relative error} propagates, and $\widetilde{\rho}(E)$
approximately has a constant relative error also it might span 
ten thousands of orders of magnitude~\cite{Langfeld:2015fua}. 

\subsection{Exponential error suppression at work: the SU(2) and SU(3)
  showcase \label{sec:expsupp} }

\begin{figure}
  \includegraphics[height=7cm]{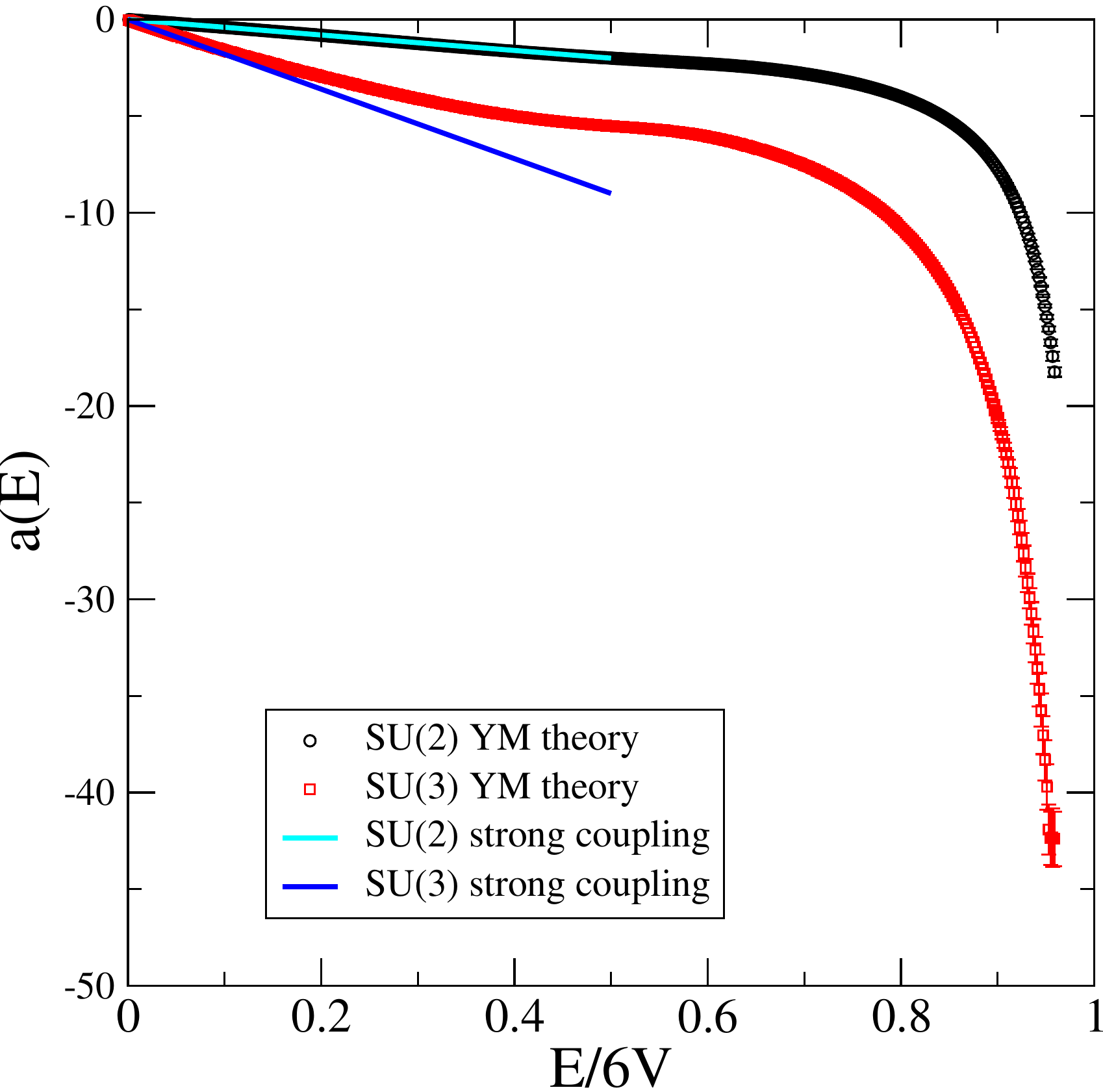}
  \hspace{0.5cm} 
  \includegraphics[height=7cm]{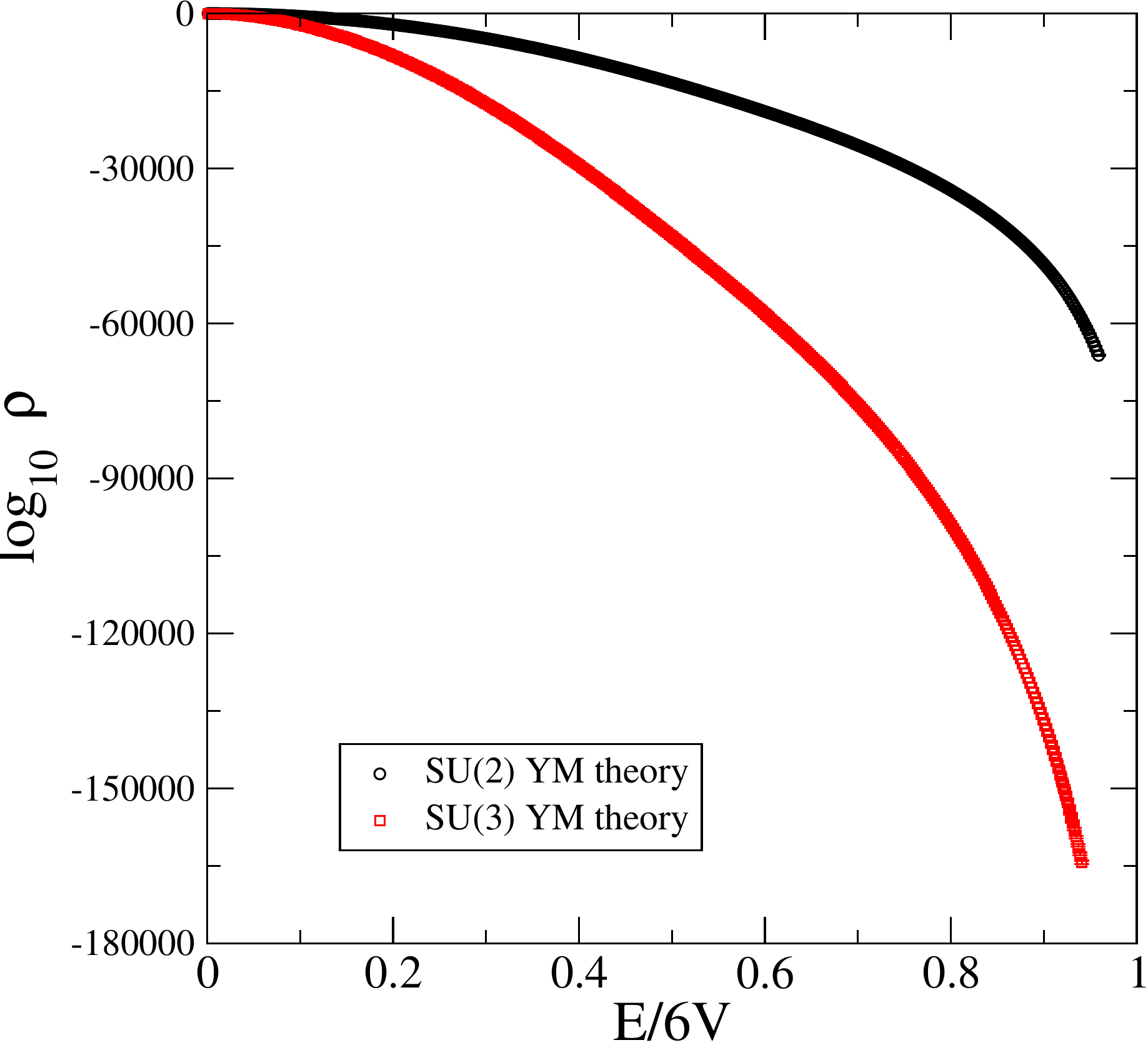}
  \caption{\label{fig:2} LLR coefficients $a(E)$ (left) and the density
    of states $\rho (E)$ (right) as a function of the action $E$ for a
  SU(2) and SU(3) lattice gauge theory on a $10^4$ lattice (plots
  from~\cite{Gattringer:2016kco}). } 
\end{figure}
For an illustration, I show results for the SU(2) and SU(3) pure
Yang-Mills theory on a $10^4$ lattice. For the LLR simulations, I used
the Gaussian window function (\ref{eq:7}) and LHMC update
algorithm. $20$ (and occasionally $40$) independent samples for $a(E)$
are obtained for carrying out the bootstrap error analysis. 

\medskip 
My numerical findings for the coefficients $a(E)$ are shown 
in the left panel of figure~\ref{fig:2}. At small 
values $a(E)$, which implies small values for $E$, $a(E)$ can be
calculated analytically using strong coupling
techniques~\cite{Gattringer:2016kco}. The result for the reconstructed
density $\rho (E)$ for SU(2) and SU(3) can be found in the right panel
of the same figure. Note that, in the SU(3) case, the density
stretches over more than $150,000$ orders of magnitude with almost a constant
{\it relative} error over the whole range.

\subsection{Ergodicity}

\begin{figure}
  \includegraphics[width=6cm]{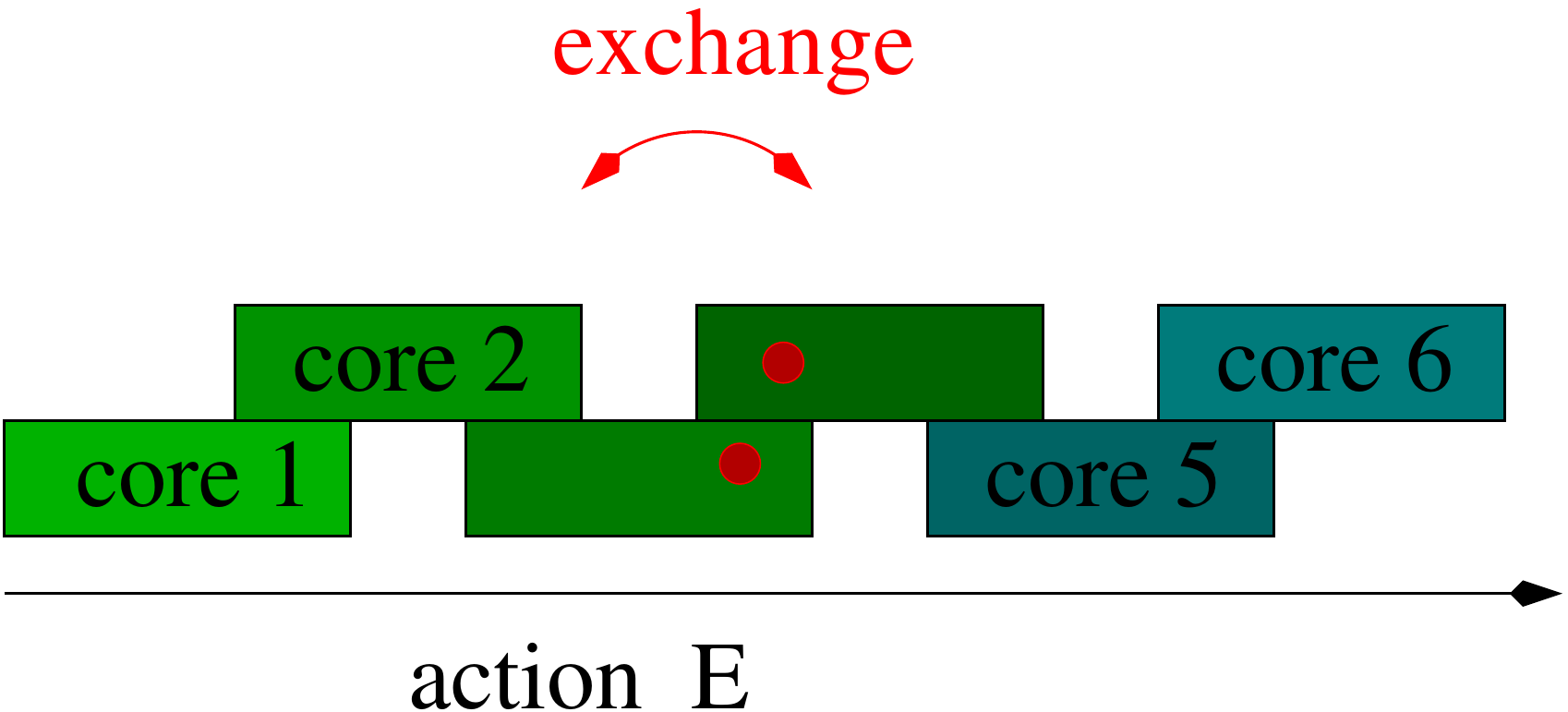}
  \hspace{1.5cm} 
  \includegraphics[width=7cm]{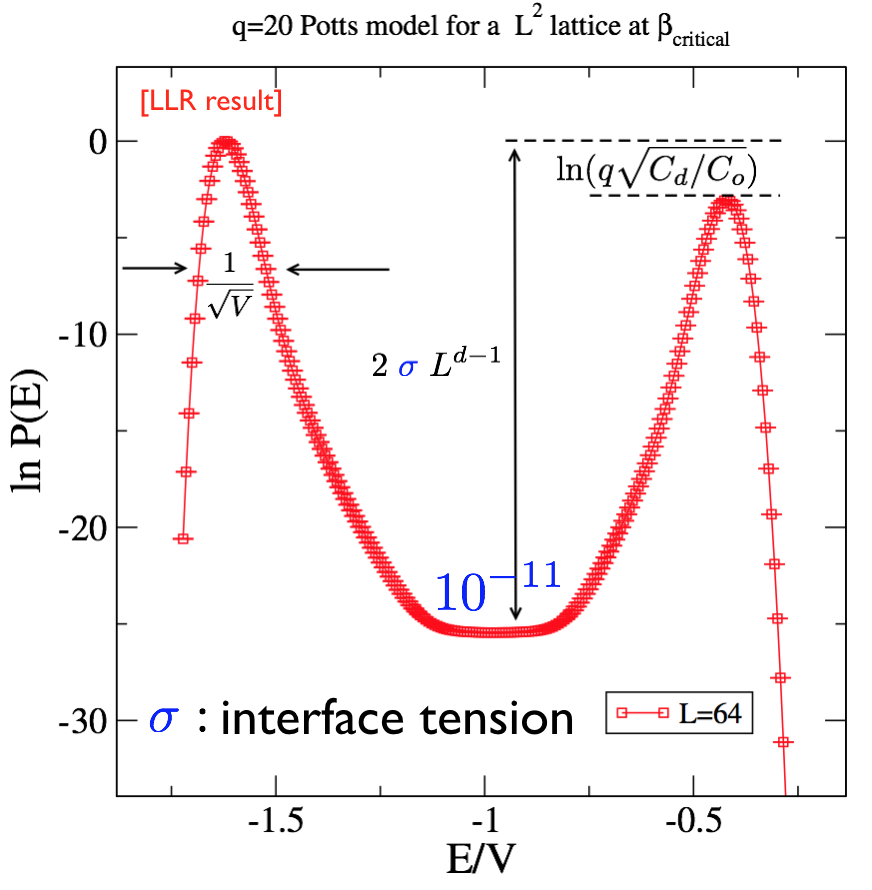}
  \caption{\label{fig:3} Left: illustration of the replica exchange
    method. Right: Typical double-peak structure of $P(E)$ of a high
    state Potts model at criticality. } 
\end{figure}
Shortly after we published the first paper~\cite{Langfeld:2012ah} on
the LLR method, concerns were raised that the approach might suffer
from ergodicity problems since the window function $\Omega _{[E,\DE]}$
might impede the exploration of the full configuration space in
particular if a step function is used for $\Omega _{[E,\DE]}$. 
To resolve this issue on both a conceptional and a practical level, we
proposed in~\cite{Langfeld:2015fua} a generalised Replica Exchange
Simulation~\cite{Swendsen1987}: The set of actions $E$ and the width
$\DE$ is chosen to produce overlapping action intervals (see
figure~\ref{fig:3} for an illustration). Each core of an HPC facility
is now performing the LLR approach and, in particular, hosts a
configuration corresponding to the action interval of the core.
If two ``neighbouring'' cores now host configurations the action of
which would be eligible for either core according to the window
function (a step-function for illustration purposes here), a
Monte-Carlo step decides whether both configurations are being
exchanged. This approach implies that a particular configuration is
not restricted to a finite action interval solving the ergodicity
problem conceptually and reduces auto-correlation time significantly in
practice.

\medskip
Let us consider the 2-dimensional q-state Potts model on a square
lattice for $q$ as high
as $20$ to illustrate the method. In two dimensions, many analytical
results are available by virtue of the exact solution of the model by
Baxter~\cite{Baxter:2000ez}. Our target will be the probabilistic
weight $P(E)$ as a function of the action $E$ at criticality. The
infinite volume critical coupling is given by
$
\beta _\mathrm{critical} \; = \; \frac{1}{2} \, \ln \, ( 1 + \sqrt{q})
\; . 
$
Figure~\ref{fig:3}, right panel, summarises the expectations: the scale
of the width of the individual peaks is set by $1/\sqrt{V} $ where
$V=L^2$ is the volume and $L$ the system size. At criticality, the
relative height is determined by $q$, the degeneracy in the broken
phase, and the ratio of the specific heats in ordered and disordered
phase. Between the peaks, a plateau emerges, which is due to the
formation of 1-dimensional interfaces. In $d$ dimension, the
expectation is that the ``probability floor'' is given by
$
\exp \{ - 2 \sigma L^{d-1} \} \; , 
$
where $\sigma $ is the interface tension. 

\subsection{Ergodicity - the $q=20$ state Potts model as
  showcase \label{sec:rep} }

\begin{figure}
  \includegraphics[width=6cm]{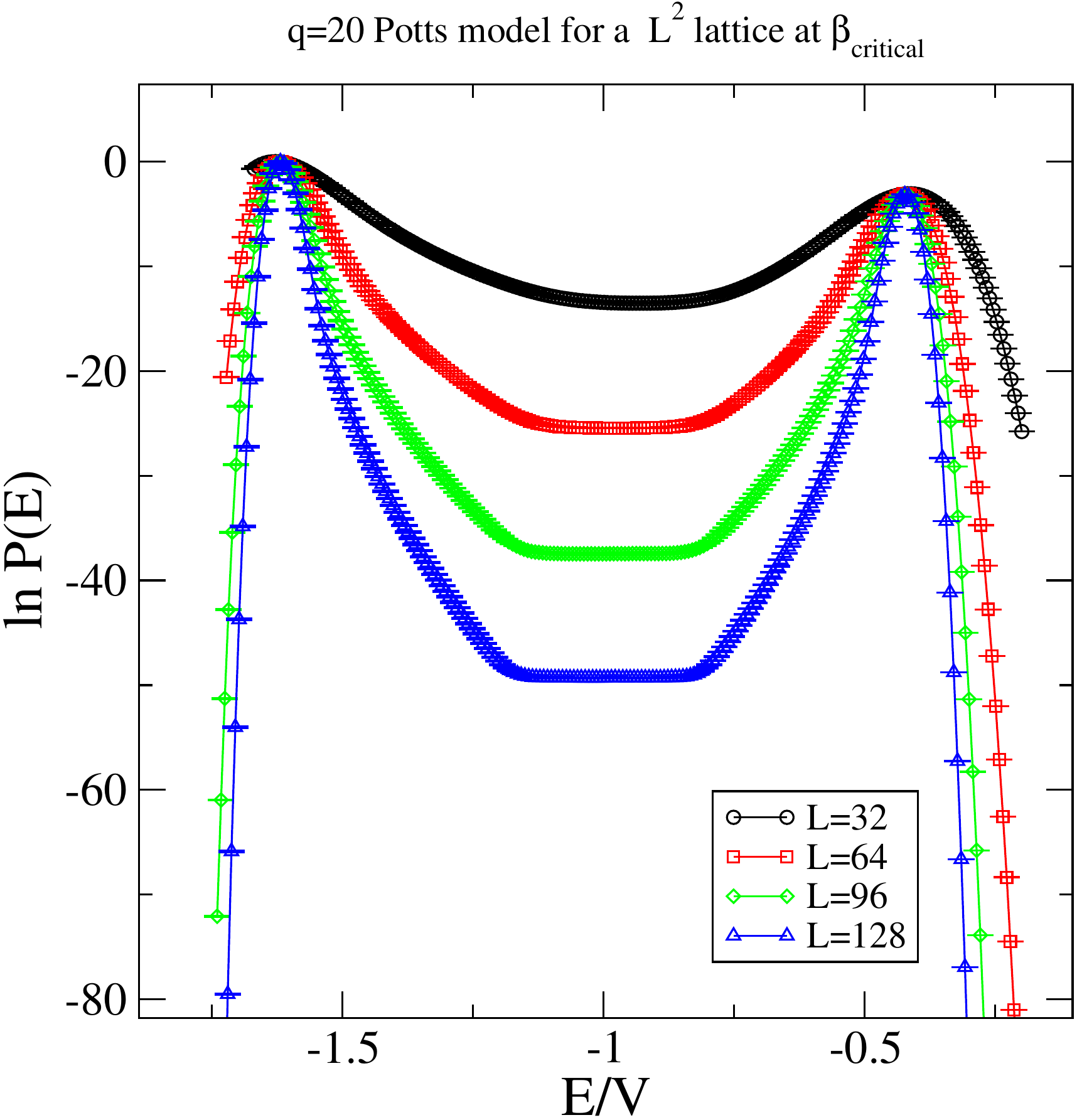} 
  \hspace{1.5cm} 
  \includegraphics[width=6.5cm]{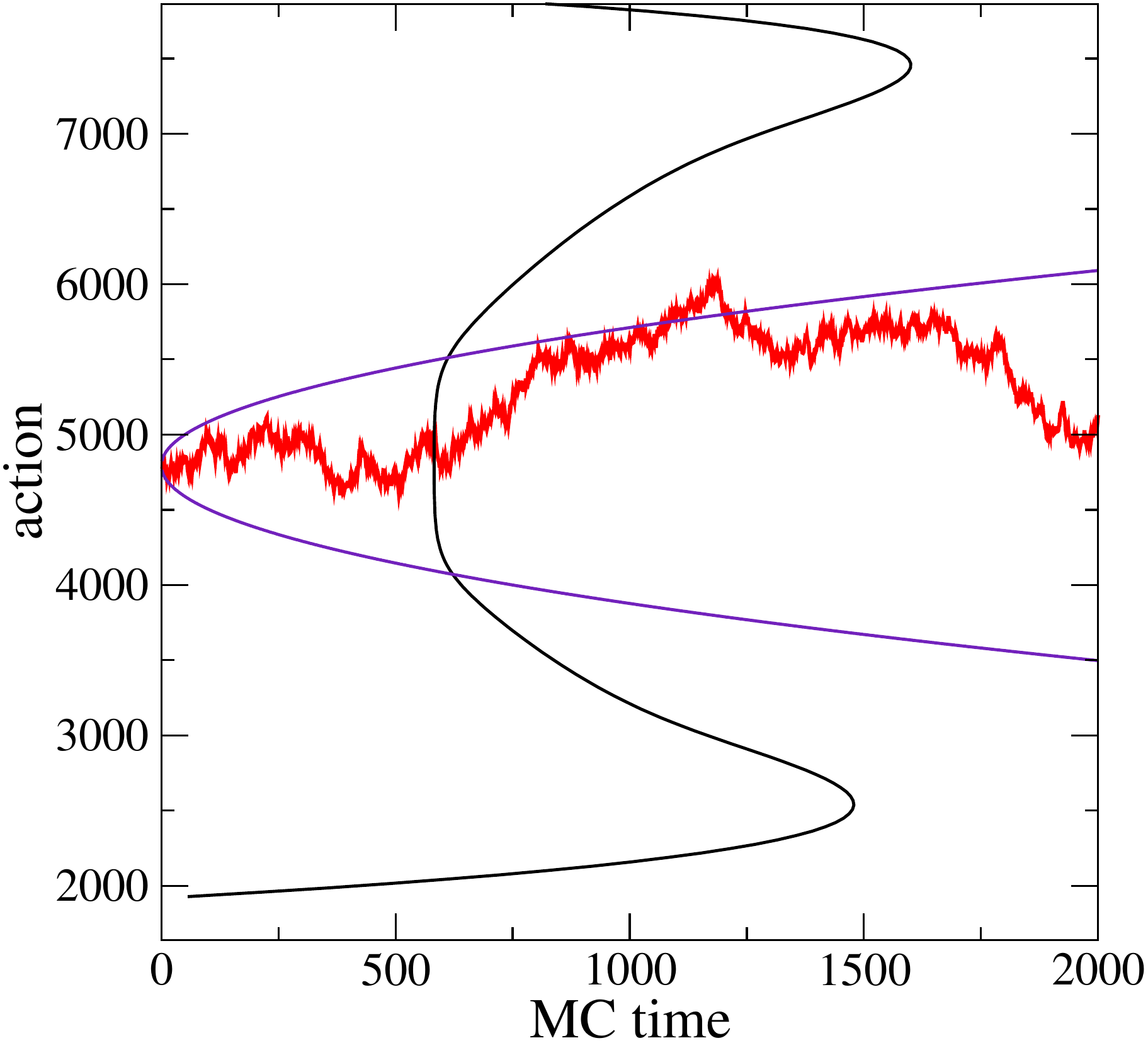}
  \caption{\label{fig:4} Left: The probabilistic weight $P(E)$ for the
  2d $q=20$ Potts model for the lattice sizes $L=32,64,96,128$ using
  the LLR method. Right: Tracing a particular configuration throughout
  the simulation: shown is the action of the configuration over
  Robins-Monro MC simulation time ($q=20,L=64$).  
  } 
\end{figure}
First numerical results for the case of $q=10$ exemplified the
Multicanonical algorithm by Berg and Neuhaus~\cite{Berg:1992qua}, and
first results for up to $q=20$ can be found in~\cite{Billoire:1993fg}.
For testing the replica exchange method, I have chosen the
step-function window 
function with action intervals that overlap by 50\%. This considerably
alleviates the  logistics of the Replica exchange step.  MC spin updates
are performed with standard local heat-bath method. Figure~\ref{fig:4},
left panel, shows my numerical result for the probabilistic weight
$P(E)$ at criticality (in infinite volume) for several values of the
lattice size $L$. Especially, the result for $L=128$ clearly shows a plateau
forming due to interfaces. Also note that, on a logarithmic scale, the
spacing between the plateau values linearly depends on the system size
$L$. This confirms the 1-dimensional nature. 

\medskip
\begin{figure}
  \includegraphics[width=7cm]{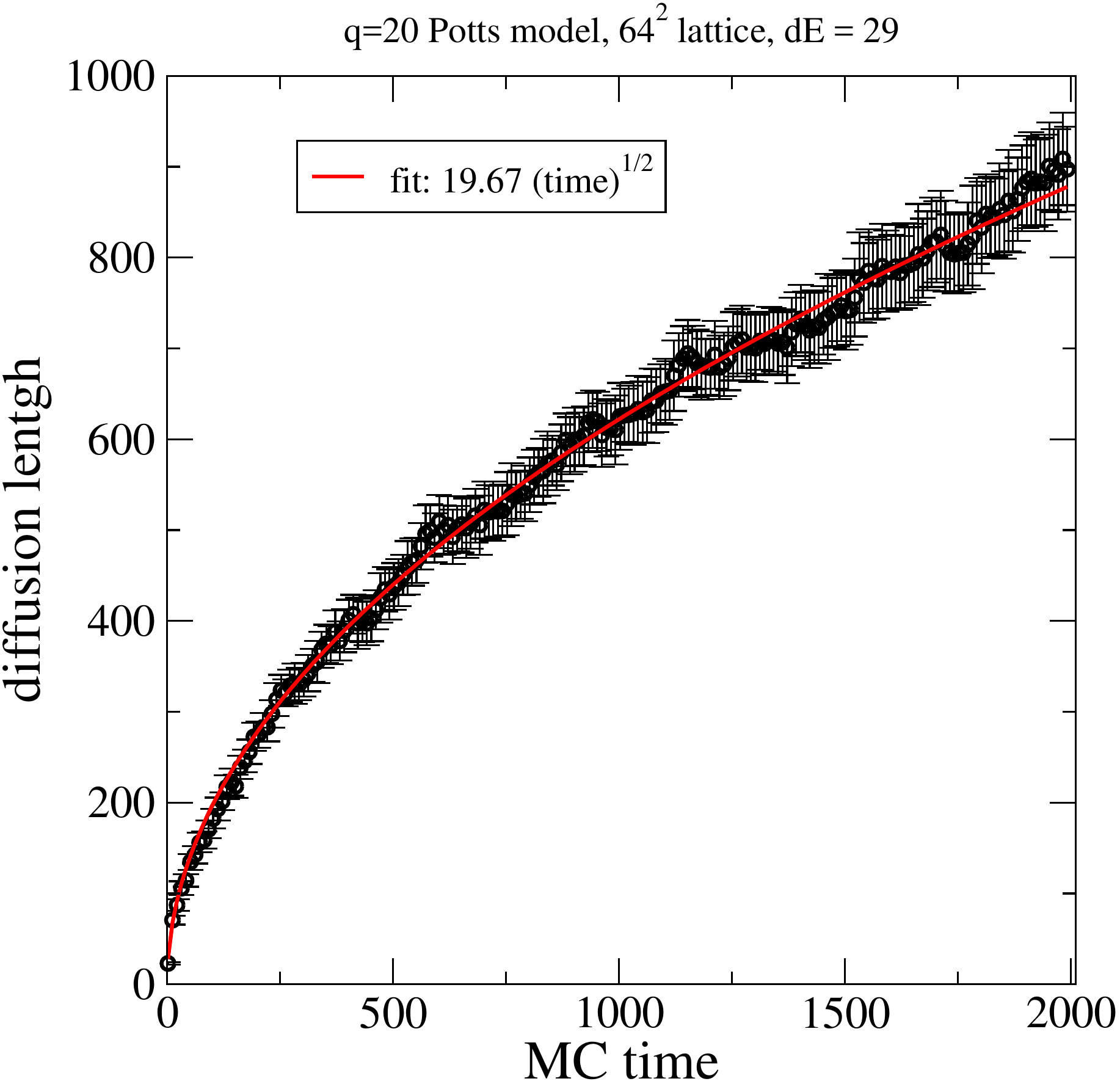} 
  \hspace{0.5cm} 
  \includegraphics[width=7.5cm]{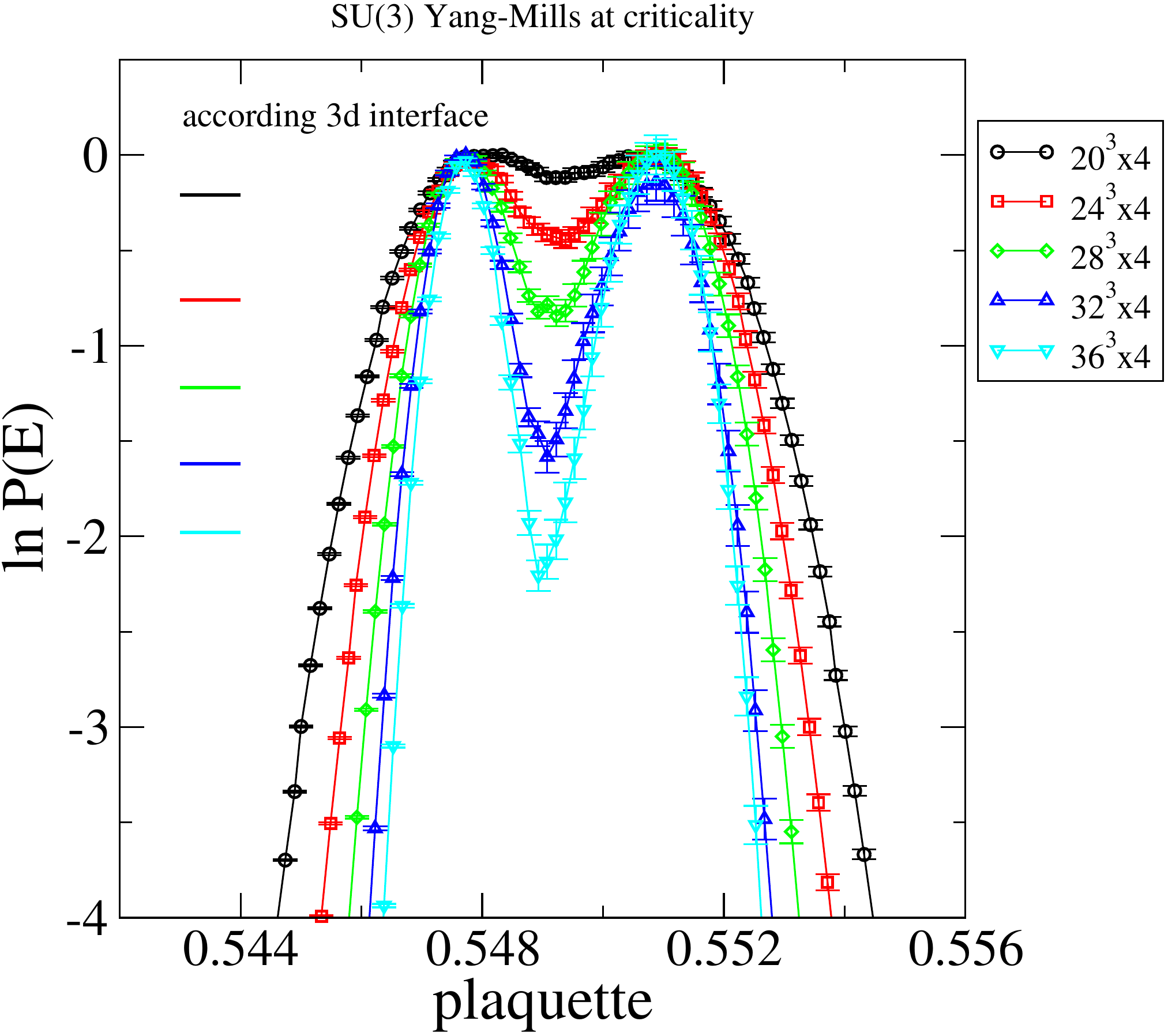}
  \caption{\label{fig:5} Left: Diffusion length as a function of the
    MC time in comparison to the expectation of a random walk
    [$20$-state Potts model at criticality and size $L=64$]. Right:
    $P(E)$ at criticality of the SU(3) Yang-Mills theory. 
  } 
\end{figure}
Once for a given action interval specified by $E$ the corresponding
coefficient $a(E)$ has (almost) converged, the MC steps do perform a
random walk in configuration space carved out by the Window function
for  the interval $[E,
  \DE]$. If Replica steps facilitate the exchange between
configurations of neighbouring action intervals, one expects that a
particular configuration drifts through all of the configuration space
unrestricted by action intervals. This is illustrated in
figure~\ref{fig:4}, right panel, for a $L=64$ lattice and for an
interval size $\DE=29$. The inlays show the probability distribution
(black) and one standard deviation for a random walk between
intervals (purple). 
I have monitored the evolution of a particular
configuration (red line) with an initial action from the interface
regime. In this particular case, we observe that this configuration
bridges $24$ action intervals within $750$ MC steps. If the 
configurations are indeed randomly swapped by the Replica Exchange
step uninhibited by any probabilistic weight, we should observe a
``diffusion length'' that grows with square root of the simulation
time. My numerical findings in figure~\ref{fig:5} are indeed very well
fitted by this square-root law thus confirming the non-Markovian
Random walk.         

\subsection{Application - SU(3) Yang-Mills theory at
  criticality \label{sec:latent} }

Let us consider the partition function of pure SU(3) Yang-Mills
theory using the Wilson action, i.e.,
$$
Z(T) \; = \; \int {\cal D}U_\mu \; \exp \Bigl\{ \beta \sum _{x, \mu >
  \nu} \frac{1}{3} \mathrm{Re} \, P_{\mu \nu}(x) \; \Bigr\} \, , 
$$
where $P_{\mu\nu}$ is the trace of the plaquette, and $U_\mu (x) \in
$SU(3) are the gluon degrees of freedom. The theory is discretised
using a lattice of size $L^3 \times N_t$ with periodic boundary
conditions. As usual, the temperature is given by
$
T \; = \; 1 / N_t a(\beta) \; . 
$
Here, we define criticality (at a finite volume), if $P(E)$ features a
double-peak structure with both peaks of {\it equal } height. 
The critical value $\beta _\mathrm{crit}$ of the Wilson coupling is
obtain by fine-tuning until equal height is achieved.
Any density-of-states method is ideally suited for this task since,
once the density $\rho (E)$ is obtained, the probabilistic weight
$P(E) = \rho (E) \exp \{\beta E\}$ is obtained for all $\beta $
values without further simulations. Once $P(E)$ is at our figure tips,
a variety of thermodynamical quantities can be obtained (see the
discussion of the Potts model and, in particular, figure~\ref{fig:3},
right panel): 
The specific heats of the confinement and the deconfinement
  (ordered) phase from the widths of the peaks; 
The latent heat from the distance between the peaks; 
The order-disorder interface tension from the volume scaling of
  the ``plateau'' between the peaks. 
In this preliminary study, I will show results for $P(E)$ for lattice
sizes 
$
N_t=4, \; L = 20,24,28,32,36 \; . 
$
My numerical findings are shown in figure~\ref{fig:5}. Even the lattice
size $36 \times 4$ is not large enough to clearly reveal the plateau
in the action range between the peaks. For large volumes, the plateau
value should scale with volume as $\ln \, P(E) \propto V^{-3}$. The
horizontal lines at the left of the graph indicate the expectation
according to the scaling law. A rough consistency could be observed, but
studies with much larger volumes are needed to confirm this.

\section{The density-of-states approach for complex action
  systems \label{sec:complex} } 

The density-of-states method is not limited to the calculation of
action observables.  For instance in~\cite{Langfeld:2013xbf} the probability
distribution of the Polyakov line was calculated with extreme
precision for a study of 2-color QCD at finite densities of heavy
quarks. Note that in the LLR approach, the only MC simulations are
with respect to the density-of-states, a quantity that is (semi-)
positive definite even if the variable of interest is not even real. 

\subsection{The generalised density-of-states}

A wide class of theories of high phenomenological interest is spanned
by quantum field theories at finite densities. The prototype partition
function is given by: 
\be 
Z(\mu ) \; = \; \int {\cal D} \phi \;  \e ^{S_R[\Phi ](\mu )} \; \exp
\Bigr( i \mu \, S_I[\phi ] \Bigl)  
\label{eq:30}
\en 
and features a complex action. In such cases, the {\it generalised
density-of-states}, i.e., 
\be 
P_\beta (s) \; = \; N \; \int {\cal D} \phi \; \delta \Bigl( s \, - \, 
S_I[\phi] \, \Bigr) \;  \e ^{S_R[\Phi ](\mu )} \; ,
\label{eq:31}
\en 
is a measure for the probability of finding a particular value $s$ for
the complex part of the action. 
An early attempt to estimate $P_\beta (s)$ was proposed by
Gocksch~\cite{Gocksch:1988iz} . If $P(s)$ is known, the partition
function is recovered by a 1-dimensional Fourier transform: 
\be 
Z(\mu ) \; = \; \int ds \; P_\beta (s) \; \exp \{ i \mu s \} \; . 
\label{eq:32}
\en 
Not that $s$ is an extensive quantity. An indicative probability
distribution and the resulting partition function is given by 
$$ 
P_\beta (s) \; = \; \exp \left\{ - \frac{s^2}{V} \right\} \; , \hbo Z
= \int ds \; \e ^{-s^2/V } \, \exp \{ i \, \mu \, s \} 
\propto \exp \left\{ - \frac{\,u ^2 }{4} \, V \right\} \; .  
$$
This illustrates the scale of the problem: the integrand is of order
one (at small $s$) and only known via numerical estimators for a realistic
theory. The result, i.e., $Z$, is typically suppressed by many orders
of magnitude for large volumes $V$. To achieve an acceptable signal to
noise ratio, it is a prerequisite for success that the density-of-states
is obtained over many orders of magnitude. This is what the LLR method
can deliver. Despite of the exponential error suppression inherent to the LLR
approach (see~\ref{sec:expsupp}), it is not clear if the achieved
precision for $P_\beta (s)$ is high enough to give meaning full
results after Fourier transform. 

\subsection{Fourier transform of the generalised density-of-states } 

\begin{figure}
  \includegraphics[height=6cm]{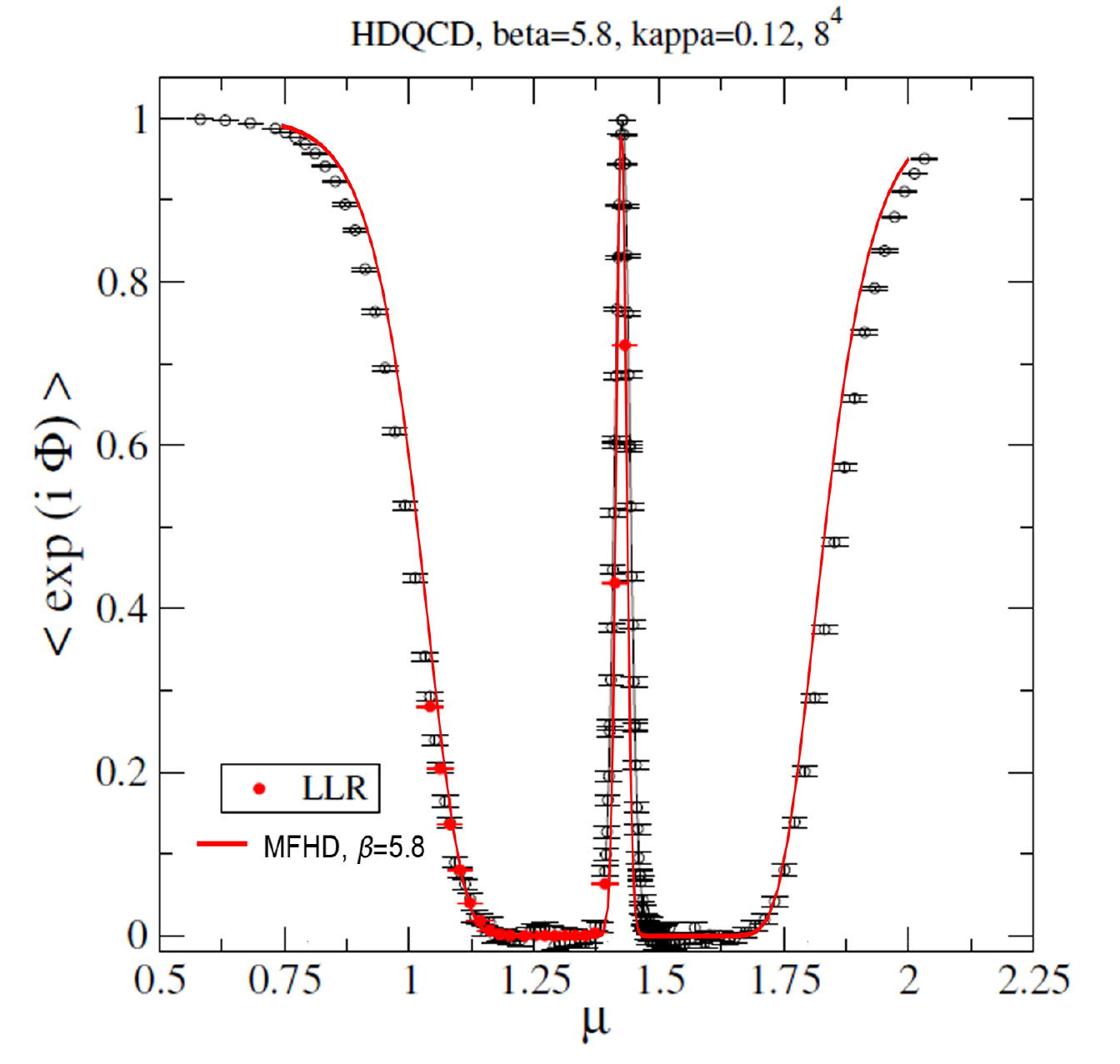} 
  \hspace{0.5cm} 
  \includegraphics[height=6cm]{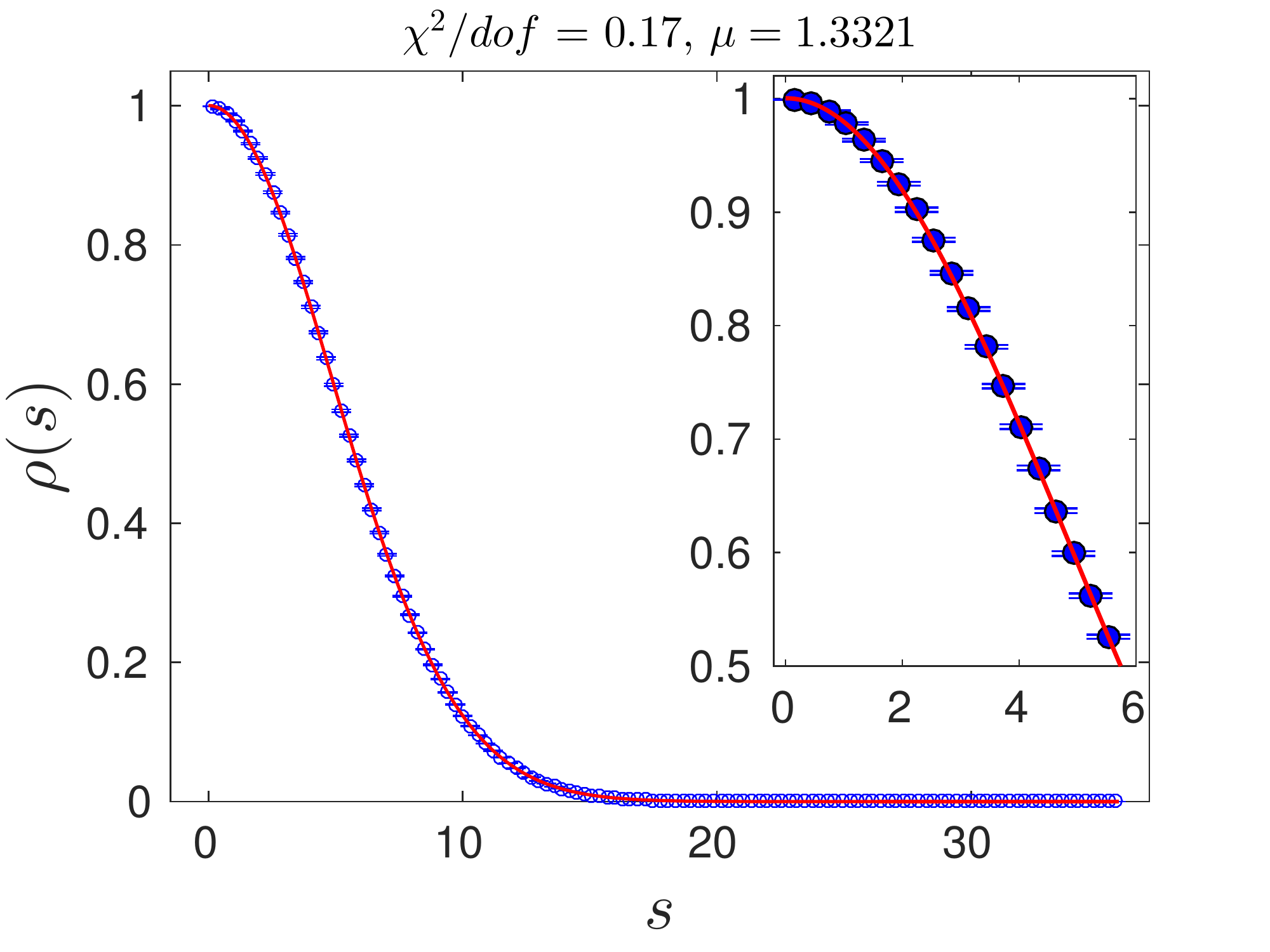}
  \caption{\label{fig:6} Left: The phase factor expectation value
    using re-weighting on the basis of a phase-quenched simulation
    (black symbols) and the mean-field result by Rindlisbacher and de
    Forcrand. Right: Probability distribution of the phase for $\mu
    =1.3321$ and the corresponding polynomial fit.
  } 
\end{figure}
Using the $Z_3$ spin model as showcase, Lucini and myself were able to
show that good results {\it can } be achieved by a Fourier
transformation of the generalised 
density-of-states~\cite{Langfeld:2014nta}. The method that we put
forward was: 

(i) Perform a polynomial fit of the logarithm of the density-of-states with
$p$ the   degree of the polynomial: $ \ln \, P_\beta (s) \; = \; \sum
_{i} ^p c_i s^i \; . $

 (ii) Calculate the Fourier transform of $\exp \{
\sum _{i} ^p c_i s^i   \}$ semi-analytical. 

 (iii)  Increase $p$ and check for stability.

\begin{figure}
  \includegraphics[height=6cm]{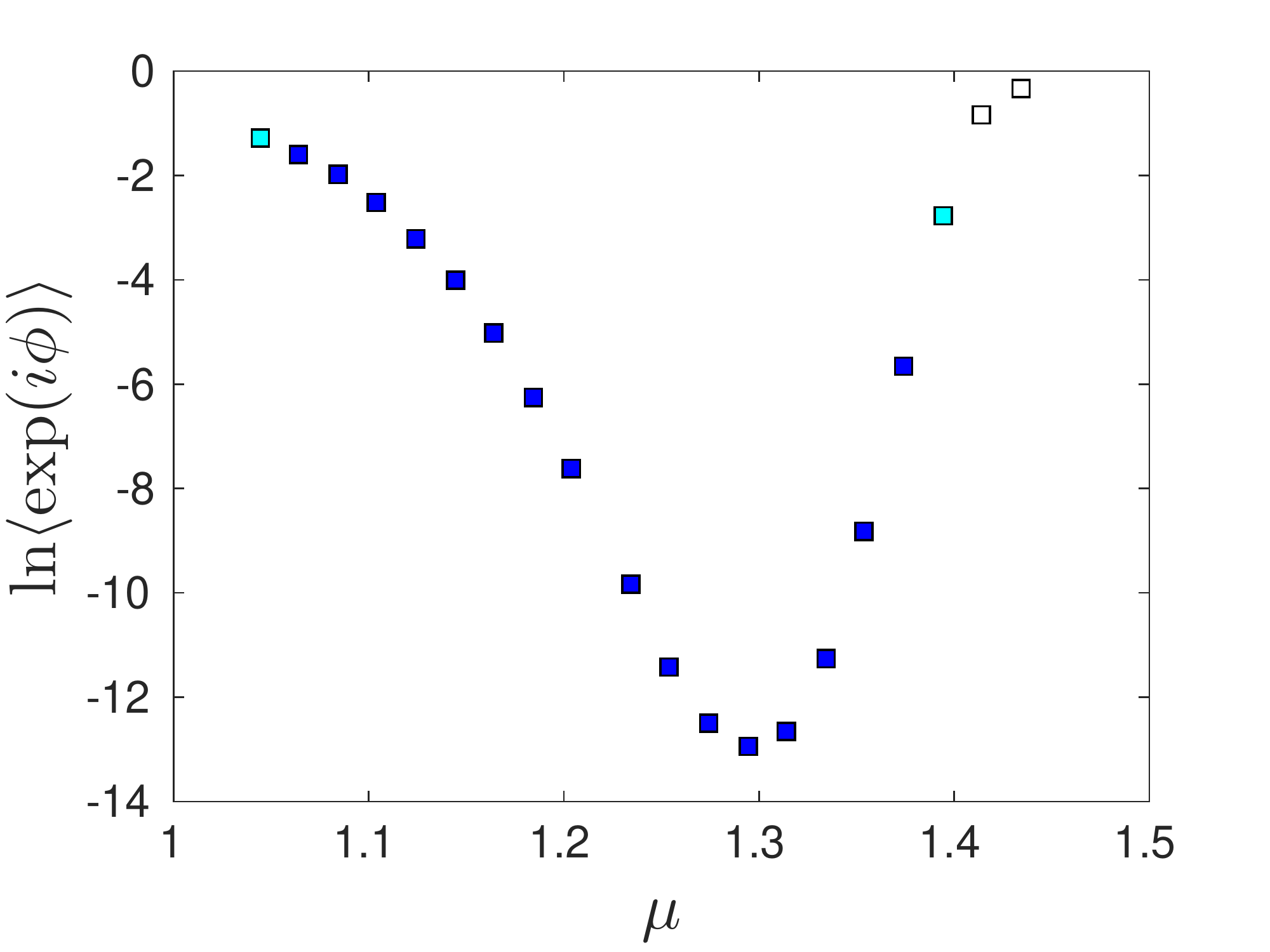}
  \includegraphics[height=6cm]{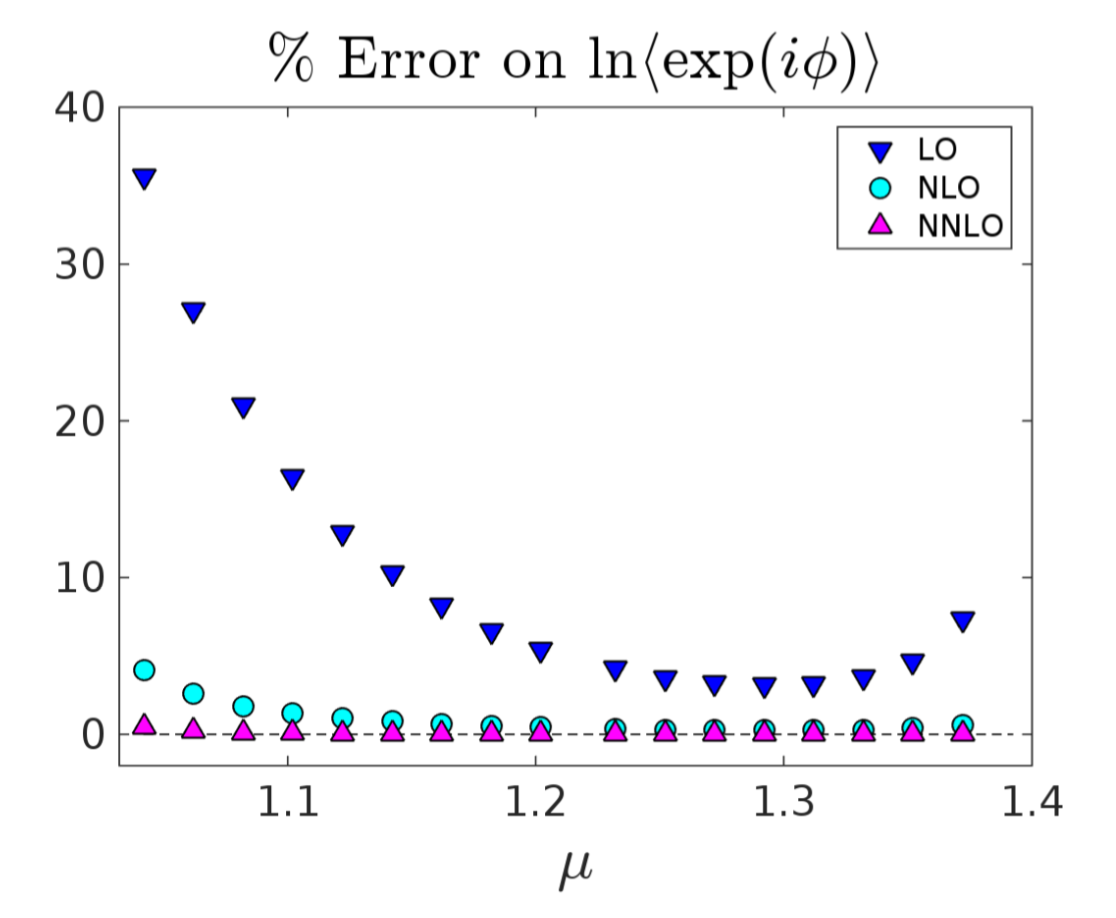}
  \caption{\label{fig:7} Left: The phase factor expectation value
    using the LLR method, polynomial fit and semi-analytic Fourier
    transformation (as outlined in the text). Right: relative error of
    the extended cumulant expansion over the whole $\mu $ range. 
  } 
\end{figure}

Recently, Garron and myself also applied the method in the context of QCD at
finite densities of heavy quarks, the so called
HDQCD (see~\cite{Garron:2016noc}). Starting point is the QCD partition function 
\be 
Z(\mu) \; = \; \int {\cal D} U_\mu \; \exp \Bigl\{ \beta
S_\mathrm{YM}[U] \Bigr\} \; \mathrm{Det} M(\mu) \; , 
\label{eq:33}
\en 
where $M(\mu)$ is the quark operator. In the so-called heavy dense
limit~\cite{Bender:1992gn,Blum:1995cb} of both, large quark mass $m$ and large
chemical potential $\mu $ with a finite ratio of $\mu /m $, the quark
determinant factorises in a product of local determinants. This
feature greatly simplifies the numerical approach leaving HDQCD as an
ideal testbed for methods dealing with sign problems (see
e.g.~\cite{Aarts:2016qrv}). A standard  MC simulation of 
the phase quenched theory and re-weighting with respect to the phase
factor of the quark determinant provides a first glimpse on the
severeness of the sign problem. Our re-weighting result is shown as a
function of the chemical potential in figure~\ref{fig:6} in comparison
with that from a mean-field
approach~\cite{Rindlisbacher:2015pea,Akerlund:2016myr}~\footnote{I am
  grateful to Tobias Rindlisbacher and Philippe de Forcrand for
  providing the mean field result.}. HDQCD possesses a variety of
interesting features: in the large mass limit, the graph exhibits a
particle-duality and therefore a symmetry around $\mu = m $. The
theory is a also real at the threshold $\mu = m$ (or at so-called
half-filling: at threshold, the density is half of the saturation
density at large $\mu $). Of high importance here is that we encounter
a strong sign problem close to threshold $\mu < m$ where the
re-weighting approach fails to produce reliable data. Using $\mu =
1.3321$, which falls into the strong-sign problem regime,
figure~\ref{fig:6} also shows the generalised probability distribution
$P_\beta (s) $ where $s$ is the sum of all the phases of the local
determinants (see~\cite{Garron:2016noc} for details). Also shown in
this graph is the polynomial fit. As for the $Z_3 $
theory~\cite{Langfeld:2014nta}, we empirically observe the importance
of {\it data compression}: whenever we succeeded to represent hundreds of
numerical data points of $\ln \, P_\beta (s)$ by a few (e.g. of order
7) fitting coefficients, the subsequent semi-analytical Fourier
transformation produces reliable and significant results with
acceptable size of the bootstrap errors. For HDQCD, figure~\ref{fig:7} shows our
final result for the logarithm of the phase factor expectation value
as a function of the chemical potential. Bootstrap error bars are
within the plotting symbols. 

\subsection{Extended cumulant expansion for complex action systems} 

The success of the method described in the previous subsection hinges
on the need to find a fit function that represents the multitude of
data with relatively few fit parameters. This approach was
successful for the $Z_3$ theory and HDQCD for the strong-sign problem
regime, but might fail if such a fit function cannot be found. Here,
we wish to propose an alternative more systematic approach to the
Fourier transform of the density-of-states. If $\phi \in ]-\pi, \pi ]$ denotes the
phase of the quark determinant, it is well known~\cite{Saito:2013vja,Greensite:2013gya}
that the probability distribution (or generalised density-of-states) is
almost uniform in the strong sign problem regime and that deviations
of the Gaussian distributions are suppressed by powers of the volume
$V$: 
\be 
P(\phi ) \; = \; 1 \; + \; \epsilon \, a_1 \; \phi ^2 \; + \; \epsilon
^3 \, a_2 \; \phi ^4 \; + \; \epsilon
^6 \, a_3 \; \phi ^6 \; \ldots \; , \hbo \epsilon = 1/V \; , 
\label{eq:40} 
\en 
where (for large volumes) the coefficient $a_i$ are volume independent
and non-zero. As usual, the phase factor expectation value is given by Fourier
transform: 
\be 
\langle \e ^{i \phi } \rangle \; = \; \frac{1}{N} \int _{-\pi }^\pi
d\phi \; P(\phi) \; \e ^{i \phi } \; , \hbo N \; = \;  \int _{-\pi }^\pi
d\phi \; P(\phi) \; . 
\label{eq:41} 
\en 
Our method~\cite{Garron:2016nrm} performs a systematic expansion of $\langle \e
^{i \phi } \rangle  $ with respect to $\epsilon $. The zeroth order
$P(\phi )=1$ trivially yields $\langle \e^{i \phi } \rangle  =0 $,
which is good starting point for the strong sign-problem
regime. Using the standard moments, i.e., 
$
\langle \phi ^n \rangle = \frac{1}{N} \int _{-\pi} ^ \pi \phi ^n \;
P(\phi) \; d \phi \; , 
$
we define the {\it extended cumulants } by 
\bea 
M_4 &=& \langle \phi^4 \rangle \; - \; \frac{ 3 \pi^2 }{5} \, \langle
\phi ^2 \rangle \; = \; {\cal O}(\epsilon ) \; , 
\hbo 
M_6 \; = \;  \langle \phi^6 \rangle \; - \; \frac{ 10 \pi^2 }{9} \, \langle
\phi^4 \rangle \; + \; \frac{ 5\pi^4 }{21} \, \langle \phi^2 \rangle
\; = \; {\cal O}(\epsilon ^3 ) \; ,  
\nonumber \\ 
M_8 &=& \langle \phi^8 \rangle \; - \; \frac{ 21 \pi^2 }{13} \, \langle
\phi^6 \rangle \; + \; \frac{ 105\pi^4 }{143} \, \langle \phi^4
\rangle \; - \; \frac{ 35 \pi^6 }{429} \, \langle
\phi^2 \rangle \; = \; {\cal O}(\epsilon ^6 ) \; . 
\nonumber 
\ena 
The systematic expansion of the phase factor expectation value then
can be facilitated by means of these moments: 
\be 
\langle \mathrm{e}^ {i \phi } \rangle = - \frac{175}{2 \pi ^6} \; M_4 \; + \;  \frac{4851 \, ( 27 - 2
  \pi^2) }{ 8 \pi^{10}} \; M_6 \; - \; \frac{57915\,  (3\pi^4
-242 \pi^2 +2145 ) }{16 \pi ^{14} } \; M_8 
\; + \; {\cal O}(\epsilon ^8) \; .
\label{eq:45} 
\en 
Garron and myself have tested the moment expansion in the context of
HDQCD. Choosing $\mu = 1.2921$ from the strong sign-problem regime, we
find order by order: 
\be 
\langle \mathrm{e}^{i \phi } \rangle \; = \; 10 ^{-6} \; \Bigl[ 
1.45(28) \; + \; 0.67(13) + 0.068(13) \Bigr] \; + \; {\cal O}
(\epsilon ^8)
\label{eq:46} 
\en 
Figure~\ref{fig:7}, right panel, shows the relative error of the
result of the extended cumulant expansion as a function of $\mu $ in
relation to  the ``exact'' results from the previous subsection. A
expected, the expansion works best in the strong sign-problem regime
where the deviation of $P(\phi )$ from uniformity are smallest. Note,
however, that even at small $\mu $, the method yields remarkable
results.

\section{Conclusions} 

Importance sampling Monte-Carlo based upon local updates fails for
systems near first order transitions due to a suppression of tunneling
between states of equal importance. They also fail if the Gibbs factor
is not (semi-) positive definite, which excludes the large class of
quantum field theories at finite matter densities. Cluster algorithms
provide a solution if one succeeds to identify and update physically
relevant clusters. The definition of physical clusters might also
emerge from dualisation, and the dual theory might be real solving the
sign problem in this case as
well~\cite{Chandrasekharan:2008gp,Langfeld:2013kno,Gattringer:2014nxa}. 

\medskip  
Rather than devising global update strategies, we can retain local
updates and abandon Importance Sampling. Non-Markovian Random Walks
propose updates with a self-improving estimate of the
density-of-states, thus incorporating memory, until a random walk in
configuration space is reached. This approach bears the potential to overcome the
ergodicity problems near first order transitions. An important attempt
is this direction is the multicanonical
algorithm~\cite{Berg:1992qua,Billoire:1993fg}. However, this algorithm
attempts an update of the density-of-states over the whole action
range at a time, and the initial guess of this density severely limits the
performance of the algorithm near criticality. Significant progress is
provided by the Wang-Landau type techniques~\cite{Wang:2001ab}: while
the update is again informed by the density-of-states, updates are
limited to action windows, and results are collated once a
random walk is achieved in each action window. Note that at this
stage, the overlap problem has been avoided. The LLR method that has
been extensively discussed in this paper belongs to this class. While
the standard Wang-Landau algorithms are based upon action histograms,
the LLR method is based upon moments of the probability
distribution avoiding histograms at all. This has advantages for
theories with continuous degrees of freedom. 

\medskip 
Subsequent to the first publication on LLR in
2012~\cite{Langfeld:2012ah}, concerns have been raised in relation to
the action window impeding ergodicity and the range of
applications. Excellent progress has been made over the recent years
and some of it has been summarised in his paper: 

\begin{itemize} 
\item Using a Replica Exchange within the LLR iteration (see
  subsection~\ref{sec:rep}) guarantees conceptual ergodicity. Moreover
  at a practical level, we have not observed any critical slowing down
  or ergodicity issues for the $q=20$ state Potts model (see
  also~\cite{LuciniLatt}). A systematic study of the volume
  scaling of thermodynamical quantities at criticality is left to
  future work. 
\item Using a smooth window function (see subsection~\ref{sec:latent})
  featuring in LLR expectation values allows the design of very efficient
  algorithms (see also~\cite{PellegriniLatt}). On the basis of LHMC, 
  results for the latent heat of a pure SU(3) Yang-Mills theory at
  criticality have  been firstly reported here  (see
  subsection~\ref{sec:latent}, see also~\cite{EjiriLatt}). 
\item The LLR approach can be used to calculate any
  observables alongside the density-of-states not just action based
  observables. The theoretical framework for this has been presented
  in~\cite{Langfeld:2015fua}. Note also that the approach can be easily
  generalised to calculate observables with unprecedented
  precision. This has been e.g.~achieved for the SU(2) Polyakov
  line~\cite{Langfeld:2013xbf}. 
\item The LLR approach is not hampered by a complex action and is
  therefore a candiate algorithm to study finite density quantum field
  theory. Here, the numerical task is to extract a very small signal
  from the Fourier transform of the (generalised) density-of-states
  (see section~\ref{sec:complex}). The LLR algorithm features the
  remarkable property of an exponential error suppression, and it has
  been shown that at least for some theories a solution of sign
  problem in the theory's original formulation is indeed
  feasible~\cite{Langfeld:2014nta,Garron:2016noc}. 
\end{itemize}

 Non-Markovian Random walks as the Wang-Landau type methods such as
 the LLR algorithm might provide the key to overcome critical slowing
 down inherent to theories near first order criticality. More
 important, however, is that the generalised LLR approach to complex
 action systems is one of the rare methods available today that are
 exact, do not rely on a specific property of the theory and that
 deliver controllable errors all at the same time. 

\bigskip 

{\bf Acknowledgements: } It is pleasure to thank my collaborators on
LLR projects: L.~Bongiovanni, Ch.~Gattringer, N.~Garron, B.~Lucini, J.~Pawlowski,
R.~Pellegrini, A.~Rago.

\end{document}